\documentclass{emulateapj}

\usepackage{amsmath, amsthm, amssymb}

\shorttitle{Turbulence from Gradients of Linear Polarization}

\shortauthors{BURKHART ET AL.}
\begin{document}

\title{Properties of Interstellar Turbulence from Gradients of Linear Radio Polarization Maps}

\author{ Blakesley Burkhart\altaffilmark{1}, A. Lazarian\altaffilmark{1} \&  B. M. Gaensler\altaffilmark{2} }
\altaffiltext{1}{Astronomy Department, University of Wisconsin, Madison, 475 N. 
Charter St., WI 53711, USA}
\altaffiltext{2}{Sydney Institute for Astronomy, School of Physics, The University of Sydney, NSW 2006, Australia}

\begin{abstract}
Faraday rotation of linearly polarized radio signals provides a very sensitive probe of fluctuations
in the interstellar magnetic field and ionized gas density resulting from magnetohydrodynamic (MHD) turbulence. We used a set of statistical tools to analyze images of the spatial gradient of linearly polarized radio emission ($|\nabla \textbf{P}|$) from the ISM for both
observational data from a test image of the Southern Galactic Plane Survey (SGPS)  
and  isothermal simulations of MHD turbulence. We compared the observational data with results of
synthetic observations obtained with the simulations of 3D turbulence. Visually, in both data sets, a complex network of filamentary structures is seen. Our analysis shows that the filaments in the gradient can be produced by shocks as well as random fluctuations characterizing the non-differentiable field of MHD turbulence. The
latter dominates for subsonic turbulence, while the former dominates for supersonic turbulence. We
show that these two regimes exhibit different distributions as well as different morphologies in the maps of  $|\nabla \textbf{P}|$. 
Particularly, filaments produced by shocks show a characteristic ``double-jump'' profile at the sites of shock fronts resulting
from delta function like increases in the density and/or magnetic field.  Filaments produced by subsonic turbulence show lower values of $|\nabla \textbf{P}|$, have a single jump profile, 
and are more dominated by fluctuations in the magnetic field.
In order to quantitatively characterize these differences we use the topology tool known as a genus curve as well as
the moments of the image distribution, i.e. the mean, variance, skewness and kurtosis.  
We find that higher values for the moments correspond to cases of $|\nabla \textbf{P}|$ with larger Mach numbers, but the strength of the dependency is 
connected to the  telescope angular resolution: higher resolution yields stronger differences between the sonic Mach number regimes.
In regards to the topology, the  supersonic filaments observed in $|\nabla \textbf{P}|$ have a positive genus shift, 
which indicates a  ``swisscheese'' like topology, while the  subsonic cases show a negative genus, indicating a ``clump'' 
like topology.  Transonic simulations have a more neutral topology.  In the case of the genus, the dependency on the telescope
resolution is not as strong, making this tool particularly useful for observational studies. 
Based on the analysis of the genus and the higher order moments, the  SGPS test region data has a distribution and morphology that matches subsonic to 
transsonic type turbulence, which independently confirms what is now expected for the WiM.
Our initial statistical application to gradients of
linear polarized radio
data shows this type of data to be very promising for studies of turbulence in the ISM in the way of obtaining the Mach numbers.  

\end{abstract}

\section{Introduction}
\label{intro}

The interstellar medium (ISM) of our Milky Way is host to a variety of physical mechanisms that regulate and govern 
the structure and evolution of the Galaxy.   The current understanding of the ISM  is that it is a multi-phase environment 
composed of a tenuous plasma, consisting of gas and dust, which is both magnetized and highly turbulent (Ferriere 2001, McKee \& Ostriker 2007).
 In particular, the awareness of turbulence 
as a dominant physical process in the ISM has only happened in the last decade (Elmegreen \& Scalo 2004).  
Turbulence plays a critical role in the areas of  star formation, magnetic reconnection, magnetic field amplification, 
nearly every transport process, cosmic ray acceleration, magnetic dynamo, and the physics in the intercluster medium, to name just a few 
(see Lazarian \& Vishniac 1999, Vishniac \& Cho 2001,  Elmegreen \& Scalo 2004, Lazarian 2006, Ballesteros-Paredes et al. 2007,
 McKee \& Ostriker 2007 and references therein).  
Additionally, turbulence has the unique ability to transfer energy over scales ranging from kiloparsecs down to the proton gyroradius.
This is critical for the ISM, as it explains how energy is distributed from large to small spacial scales in the Galaxy.  

\begin{figure*}
\center
\includegraphics[scale=.5]{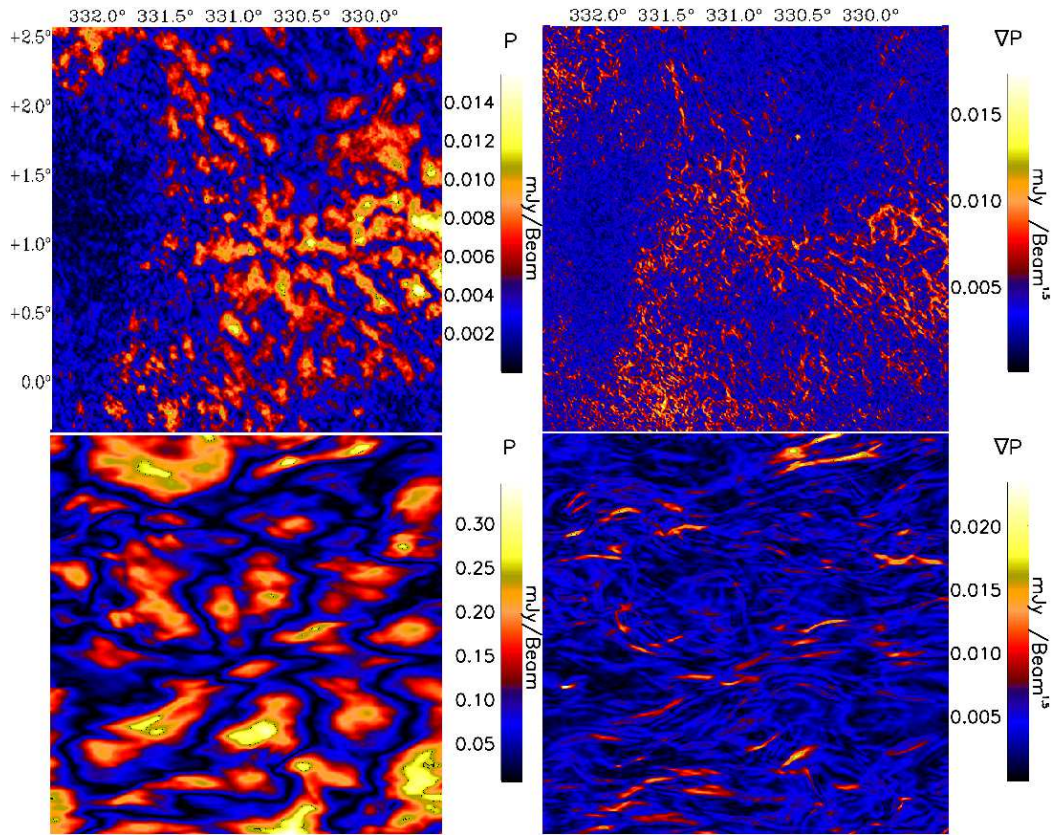}
\caption{Top: Observational P (left) and $|\nabla \textbf{P}|$ (right) from the SGPS data used in this study.
 Bottom: MHD simulation model number 1 in Table \ref{tab:models} with  P (left) and $|\nabla \textbf{P}|$ (right). 
The simulated  map of P has had the mean subtracted from maps of Q and U, similar to the observations. }
\label{fig:RM}
\end{figure*}

Despite the now obvious importance of magnetized turbulence, the situation of understanding ISM physics is no less complicated. 
In spite of the big recent advances in understanding of incompressible and compressible MHD turbulence 
(see Goldreich \& Sridhar 1995, Cho \& Vishniac 2000, Maron \& Goldreich 2001,  Cho, Lazarian \& Vishniac 2002, 2003, 
Cho \& Lazarian 2003, Kowal \& Lazarian 2010, Beresnyak \& Lazarian 2010, Beresnyak 2011) the ISM presents a complex
environment with multiple energy injection sources, different phases and various instabilities acting at different scales. The properties of this turbulence 
affect  key astrophysical processes and obtaining properties of ISM turbulence from observations opens ways of gauging numerical simulations and testing theory.
In light of these complexities, the most fruitful way of studying astrophysical MHD turbulence and the processes it affects
is to use a \textit{synergetic} approach which combines the knowledge of theoretical predictions, numerical studies, and observational efforts.

Observationally there are several ways of studying MHD turbulence. Many of these techniques hinge on 
density fluctuations in ionized or neutral media  (see Spangler \& Gwinn 1990, Armstrong et al. 1995,
Padoan et al. 2003, Falgarone et al. 2005, Chepurnov \& Lazarian 2010) and are aimed at finding the density
power spectrum of turbulence. More recently, ways to find the magnetization and the Mach number
of turbulence have been explored (Kowal, Lazarian \& Beresnyak 2007, Burkhart et al. 2009, 2010, Esquivel \& Lazarian 2010, Tofflemire, Burkhart \& Lazarian 2011, 
Esquivel \& Lazarian 2011). While column density data are 
arguable the most common type, they do not contain the full 3D picture and are only passive tracers of the turbulence velocity field. 
Various ways to study the interstellar velocity field have been explored. The
tested and theoretically motivated ways to study turbulent velocities in supersonic interstellar turbulence are based on the use of the Velocity Channel Analysis
 (VCA) and Velocity Coordinate Spectrum techniques (Lazarian \& Pogosyan 2000, 20004, 2006, 2008). These techniques have been used both for atomic HI and molecular
 data (see Stanimirovic \& Lazarian 2001, Padoan et al. 2006, 2009, Chepurnov et al. 2010) to find the spectra of the turbulent velocity fields. 
 

Naturally, studies of turbulence and magnetic fields are of great importance and synergetic value for the ISM.
 In this case, a number of techniques have been explored, e.g.
structure functions of the polarization vectors arising
from dust polarized emission (see Falceta-Gon\c{c}alves et al. 2009, Houde et al. 2011).The quantitative study of
 synchrotron intensity fluctuations can be traced to works by 
Getmansev (1958), while fluctuations of synchrotron polarization\footnote{For a theoretical description
of synchrotron fluctuations for arbitrary index of cosmic rays and realistic models of anisotropic turbulence see Lazarian \& Pogosyan (2011).}
were used, for instance, to evaluate the spectra of magnetic turbulence in Hydra cluster (En{\ss}lin \& Vogt 2006, En{\ss}lin et al.  2010)    

More recently, several authors have discussed the prospects of the use of radio polarization maps to study 
turbulence (see 
Haverkorn \& Heitsch 2004, Fletcher \& Shukurov 2006, 2007, Gaensler et al. 2011). 
Faraday
rotation maps of linearly polarized radio signals are especially promising as they provide
very sensitive probes of fluctuations in magnetic field and ionized gas density
(see Gray et al. 1998, Gaensler et al. 2001, and Landecker et al. 2010).  
The Faraday rotation can be calculated as:
\begin{equation}
 RM= K \int_l^0 n_e(l) \bf{B}(l) d{\bf l}
\end{equation}
\label{RM}

With units of rad m$^{-2}$, $K=0.81$ rad m$^{-2}$ pc$^{-1}$ cm$^{3}$ $\mu G^{-1}$ and $B$, $l$ and  $n_e$ are the magnetic field strength in $\mu G $, the 
distance along the LOS
in parsecs, and the electron density in $cm^{-3}$ along the LOS, respectively.

Although many objects seen in the polarization/Faraday maps can be
matched to objects seen in other wavelengths (such as supernova remnants), an extended diffuse polarization emission network that is rich in structure
is also present that can not be mirrored in other wavebands or in total intensity (Fletcher \& Shukurov 2007). 
The intensity variations seen in maps of Q, U and P are the result of small-scale angular structure in the
Faraday rotation induced by foreground ionized gas and are thus an indirect representation of
turbulent fluctuations in free electron density and magnetic field throughout the ISM.


In this paper we use polarization gradients to study ISM turbulence.  
The use of gradients to highlight small rapid fluctuations seen in polarization maps was first discussed
by Gaensler et al. (2011). When the spatial gradient is applied to maps of vector $\textbf{P}= (Q,U)$ a complex web of filamentary
structures is revealed.  These filaments (see right column of  Figure \ref{fig:RM} for an example)
 were interpreted by Gaensler et al. (2011) as rapid fluctuations in $n_e$ and B along the LOS 
due to turbulence.  In this paper we will further explain the origins of the filamentary structures as they are related
to turbulence and develop quantitative methods that can be applied to this data in order to obtain the Mach numbers of turbulence.
Taking the gradient of rotation measure or linear polarization maps has its advantages and disadvantages.
The primary advantage is that the spatial gradient of $\textbf{P}$ satisfies the property that it has both translational and rotational invariance
in the (Q,U) plane.  Quantities such as the polarization amplitude and polarization angle are not preserved under
arbitrary translations and rotations, which can result from one or more of a smooth
distribution of intervening polarized emission, a smooth uniform screen of foreground Faraday
rotation, or the effects of missing large-scale structure in an interferometric data-set. Thus the 
magnitude of the gradient is the simplest quantity that is not significantly effected by missing large-scale structure
or excess foreground emission or Faraday rotation.  
Taking the gradient allows one to clearly see jumps and discontinuities,
regardless of whether single-dish (total power) measurements are present in the data.  
In particular, this will highlight
areas where a sharp change in $n_e$ or $B$ occurs, which is most likely due to turbulent fluctuations or shock fronts in the ISM.  
However, one must also keep in mind the disadvantages of using gradients, namely the fact that the gradient may enhance noise.

\begin{figure}[tbh]
\centering
\includegraphics[scale=.5]{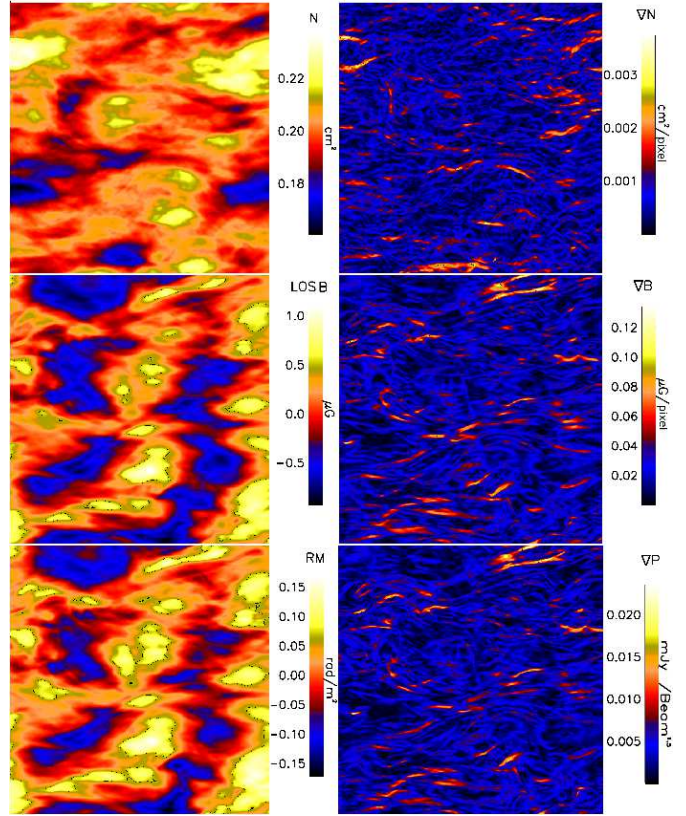}
\caption{Examples of LOS maps and their respective gradients relevant to this paper for subsonic turbulence (model 1). 
The first column shows column density, LOS magnetic field, and the rotation measure from top to bottom. 
The second column shows the gradients of column density, LOS magnetic field, and polarization vector.}
\label{fig:RM1}
\end{figure}

In this work we explore the physical causes of the filaments by taking the gradient of polarization maps of isothermal MHD turbulence.
We investigate the dependency of the sonic Mach number on the structures seen in $|\nabla \textbf{P}|$  
and calculate measures of the Probability Distribution Functions (PDFs) as well as the 
a measure of the topology called genus. Genus has been extensively used in cosmology studies (Gott et al. 1986)
and has been suggested for ISM studies in Lazarian (1999). Its use for 
synthetic column density maps and observations is discussed in the literature 
(Lazarian, Pogosyan \& Esquivel 2002, Kowal, Lazarian \& Beresnyak 2007, Kim \& Park 2007, Chepurnov et al. 2008). 

The PDFs and genus have been studied on MHD turbulence in the past and 
have both shown sensitivity to the sonic Mach number in density, column density and position-postion-velocity data (Padoan et al. 1999, Kowal, Lazarian \& Beresnyak 2007, 
Chepurnov et al. 2008, Burkhart et al 2009, 2010, Tofflemire et al. 2011).  We view these statistics as part of a set of tools which can be applied to 
polarization data in order to determine the sonic Mach number of the turbulence in the ionized ISM. We stress that the use of the whole set provides a synergetic quantity: obtaining the same result with
different techniques substantially increases how trustworthy the result is. 

The paper is organized as follows.  In \S~\ref{data} we further describe the data sets used, in particular the 
 Southern Galactic Plane Survey and a set of ideal MHD simulations, and our calculation of the rotation measure, the linear polarization maps and their gradients.
In \S~\ref{origin} we discuss the origins of the observed filaments as they relate to the sonic Mach number. 
We describe different statistical measures of the sonic Mach number in 
\S~\ref{ms}; in particular the genus and  PDFs.  
In  we discuss our results followed by conclusions in \S~\ref{concl}.

\section{Data and Method}
\label{data}
\subsection{Gradient Technique and relation of $|\nabla \textbf{P}|$ to $|\nabla RM|$}
An important observational quantity connected to interstellar density and magnetic field fluctuations is the Faraday effect. In the presence of magnetic fields
and free electrons, bifringence of circularly polarized orthogonal modes occurs, giving these modes two different propagation velocities. 
In the case of pure polarized background emission propagation through a magnetoionized medium, the linearly polarized
radiation will emerge with its polarization position angle rotated by the amount given in Equation \ref{RM}. 
Thus the relation between the observed position angle, emitted position angle, and the rotation measure is:
\begin{equation}
 \Theta-\Theta_0 = RM \lambda^2
\end{equation}
which has units of radians.  Here $\lambda$ is the wavelength in meters.

\begin{figure}[tbh]
\centering
\includegraphics[scale=.5]{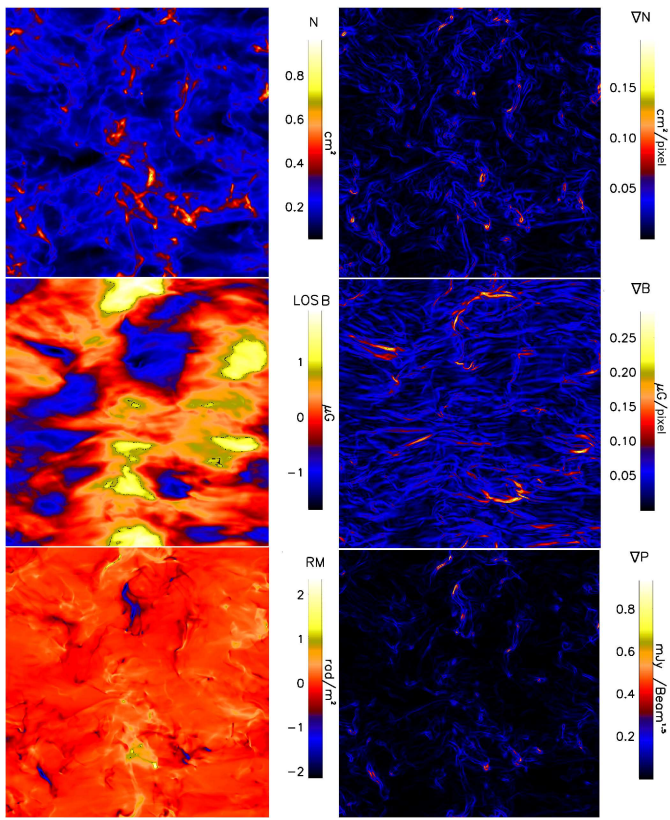}
\caption{Examples of LOS maps  and their respective gradients relevant to this paper for supersonic turbulence (model 6). 
The first column shows column density, LOS magnetic field, and the rotation measure from top to bottom. 
The second column shows the gradients of column density, LOS magnetic field, and polarization vector.}
\label{fig:RM2}
\end{figure}
Observational determination of Faraday rotation comes from measurements of the linear polarization vector $\textbf{P} \equiv$ (Q,U)
(which depends on  Stokes U and Q as $|P|=\sqrt{Q^2+U^2}$) as a function of $\lambda^2$.   
To avoid confusion between vector and scalar P, we use bold notation to denote the vector quantity of the linear polarization map.

We define the gradient of the polarization vector as:
\begin{equation}
 |\nabla \textbf{P}|=\sqrt{\left(\frac{\partial Q}{\partial x}\right)^2+\left(\frac{\partial Q}{\partial y}\right)^2+\left(\frac{\partial U}{\partial x}\right)^2
+\left(\frac{\partial U}{\partial y}\right)^2}
\end{equation}
\label{eq:grad}

We note that in the case of  vector $\textbf{P}$  we have:
\begin{equation}
 \textbf{P} = |P_0| e^{2i(RM\lambda^2+\theta_0)}
 \end{equation}
 \label{eq:exp}
   From this equation, one can derive a relationship between
 $|\nabla \textbf{P}|$ and  $|\nabla RM|$ for Faraday-thin polarized emission as:
\begin{equation}
|\nabla RM| = |\nabla \textbf{P}| / 2i \lambda ^2 |\textbf{P}|
 \label{eq:1}
 \end{equation}

Equation 5 only holds for  data in which the entire signal is measured (i.e. single-dish data included) and for which the background is uniform. 
When  $|\textbf{P}|=1$ and $\lambda =1$ (which are the assumptions we use for the simulations), one finds a trivial relation between $|\nabla \textbf{P}|$ and $|\nabla RM|$ as $|\nabla RM|= |\nabla \textbf{P}|/2i$. 
However, this relation can only be used to calculate $|\nabla \textbf{P}|$ or $|\nabla RM|$ in the simulations, since the assumption of  $|\textbf{P}|=1$ 
is almost always to simplistic for the observations because the data are missing single-dish information and/or the background $|\nabla \textbf{P}|$ is not zero.  

We calculate maps of RM, $\textbf{P}$ and their gradients from density and LOS magnetic field that is perpendicular and parallel to the mean magnetic field in the simulations.
We calculate the RM as per Equation \ref{RM} at every point and then take its spatial gradient, that is, we compute the gradient vector at every pixel of the image using
neighbor pixels. We can calculate $|\textbf{P}|$ by  calculating the stokes
vectors as $Q=\cos(2\theta)$ $U=\sin(2\theta)$.  These expressions come from applying Equation 2
 with assumed values for $\lambda$ and $\theta_0$.

We show a subsonic and supersonic case of density ($n$), $\nabla n$, LOS magnetic field (LOS B),  $\nabla B$,
RM and $|\nabla P|$ in Figure \ref{fig:RM1} and Figure \ref{fig:RM2}, respectively.  A comparison of the SGPS test data and a subsonic case is given in
Figure \ref{fig:RM}. Inspection of  maps of $|\nabla \textbf{P}|$ reveals that filaments are created 
in both cases and that there is some correlation between gradients of column density, magnetic field, and $|\nabla \textbf{P}|$.  We will discuss these further
in Section \ref{origin}.


\subsection{Southern Galactic Plane Survey }
We use a subsection of radio continuum images of an 18-square-degree patch of the Galactic plane, observed with the Australia Telescope Compact Array
(ATCA, see McClure-Griffiths et al. 2001 and Gaensler et al. 2001 for more details).
We examine the 1.4 GHz frequency data averaged over adjoining frequency channels
with simultaneously recorded Stokes \textit{I}, Stokes \textit{Q}, and Stokes \textit{U} 
as part of the SGPS test region (Gaensler et al. 2001).                                                             
This field consists of 190 mosaicked pointings of the
Australia Telescope Compact Array (ATCA) and covers the range $325.5 < l < 332.5, -0.5 < b < 3.5$. 
Complicated extended structure is seen in linear polarization throughout the test region, almost all of which has no correlation with total intensity. 
We select a $512\times512$ pixel subregion from this data to match the resolution of the simulations used in our study and display 
it in Figure \ref{fig:RM} in the top row.
The SGPS region we select begins at coordinate l=332.3373, b= -0.3138 and is not overly contaminated by bad pixels and contains significant emission.


\subsection{Simulations}
We generate a database of 3D numerical simulations of isothermal compressible (MHD)
turbulence by using the MHD code of Cho \& Lazarian (2003) and varying the input 
values for the sonic and Alfv\'enic Mach number.
The sonic Mach numbers are defined as ${\cal M}_s \equiv \langle |{\bf v}|/C_s \rangle$, 
where is ${\bf v}$ is the local velocity, $C_s$ is the sound speed, 
and the averaging is done over the whole box.
Similarly, the Alfv\'enic Mach number
is ${\cal M}_A\equiv \langle |{\bf v}|/v_A \rangle$, where 
$v_A = |{\bf B}|/\sqrt{\rho}$ is the Alfv\'enic velocity,
${\bf B}$ is magnetic field and $\rho$ is density.
 We briefly outline the major points of the numerical setup (for more details see Cho \& Lazarian (2003).

The code is a second-order-accurate hybrid essentially 
nonoscillatory (ENO) scheme (Cho \& Lazarian 2003) which solves
the ideal MHD equations in a periodic box:
\begin{eqnarray}
 \frac{\partial \rho}{\partial t} + \nabla \cdot (\rho {\bf v}) = 0, \\
 \frac{\partial \rho {\bf v}}{\partial t} + \nabla \cdot \left[ \rho {\bf v} {\bf v} + \left( p + \frac{B^2}{8 \pi} \right) {\bf I} - \frac{1}{4 \pi}{\bf B}{\bf B} \right] = {\bf f},  \\
 \frac{\partial {\bf B}}{\partial t} - \nabla \times ({\bf v} \times{\bf B}) = 0,
\end{eqnarray}
with zero-divergence condition $\nabla \cdot {\bf B} = 0$, 
and an isothermal equation of state $p = C_s^2 \rho$, where 
$p$ is the gas pressure. 
On the right-hand side, the source term $\bf{f}$ is a random 
large-scale  solenoidal driving force\footnote{${\bf f}= \rho d{\bf v}/dt$}.   

The magnetic field consists of the uniform background field and a 
fluctuating field: ${\bf B}= {\bf B}_\mathrm{ext} + {\bf b}$. Initially ${\bf b}=0$.

We scale the simulations to physical units, adopting typical
parameters for warm ionized gas.
We assume a pixel size of 0.15 parsecs and density of 0.1 cm$^{-3}$.
The simulations are assumed to be fully ionized and we do not include the effects of 
partial ionization.

To make the maps of  $|\nabla \textbf{P}|$ we first calculate the LOS rotation measure at each pixel
then we  take the take the gradient of this rotation measure map and convert it to $|\nabla \textbf{P}|$ via 
Equation \ref{eq:1}.  An equally valid way is to  calculate the rotation measure at each pixel, then shine a polarized signal through the cube 
with our assumed background values of  Q = 1, U = 0.  Then one can calculate the emergent values of Q and U by applying the simulated rotation measure map
 and then calculate the resulting $|\nabla \textbf{P}|$.  
Both methods will produce identical maps of $|\nabla \textbf{P}|$. Additional smoothing of the maps of Q and U  using a Gaussian kernel can also be 
performed to mimic the telescope beam.

\begin{table}
\begin{center}
\caption{Description of the simulations - MHD, 512$^3$
\label{tab:models}}
\begin{tabular}{cccccc}
\hline\hline
Model & $p_{gas}$ & $B_{\rm ext}$ & ${\cal M}_s$ & ${\cal M}_A$ &Description \\
\tableline
1 &2.00 &1.00 &0.5 &0.7 & subsonic \& sub-Alfv\'enic \\
2 &0.70 &1.00 &1.0 &0.7 & transsonic \& sub-Alfv\'enic \\
3 &0.10 &1.00 &2.0 &0.7 & transsonic \& sub-Alfv\'enic \\
4 &0.05 &1.00 &3.0&0.7 & supersonic \& sub-Alfv\'enic \\
5 &0.025 &1.00 &4.4 &0.7 & supersonic \& sub-Alfv\'enic \\
6 &0.0077 &1.00 &8.0 &0.7 & supersonic \& sub-Alfv\'enic \\
7 &0.0049 &1.00 &10 &0.7 & supersonic \& sub-Alfv\'enic \\
8 &2.00 &0.10 &0.5 &2.0 & subsonic \& super-Alfv\'enic \\
9 &0.70 &0.10 &1.0 &2.0 & transsonic \& super-Alfv\'enic \\
10 &0.10 &0.10 &2.0 &2.0 & transsonic \& super-Alfv\'enic \\
11 &0.05 &0.10 &3.0 &2.0 & supersonic \& super-Alfv\'enic \\
12 &0.025 &0.10 &4.4 &2.0 & supersonic \& super-Alfv\'enic \\
13 &0.0077 &0.10 &8.0 &2.0 & supersonic \& super-Alfv\'enic \\
14 &0.0049 &0.10 &10 &2.0 & supersonic \& super-Alfv\'enic \\
\hline\hline
\end{tabular}
\end{center}
\end{table}

\section{The Origin of Filamentary Structures in Polarization and Rotation Measure Gradients}
\label{origin}
A filament traced by $|\nabla \textbf{P}|$ or $|\nabla RM|$ will form as a result of  a localized change in either density or magnetic field as a function of position on the sky
as per Equation \ref{RM}.
There are many physical processes that can result in sharp changes in these quantities in the ISM, including gravitational collapse and outflows.
However, high Reynolds fluids in the ISM are expected to be turbulent (see Cho, Lazarian \& Vishniac 2003, Elmegreen \& Scalo 2004, Lazarian et al. 2009) and a more ubiquitous process responsible for fluctuations in $n_e$ or B is due to MHD turbulence, precisely because it expected  everywhere in the 
ISM\footnote{Because of the large injection scale of the ISM, the Reynolds numbers can typically reach $10^{10}$}, although the
type of turbulent environment can vary (e.g. the compressibility, magnetization, equation of state, etc.).  

\begin{figure}[h]
\centering
\includegraphics[scale=.28]{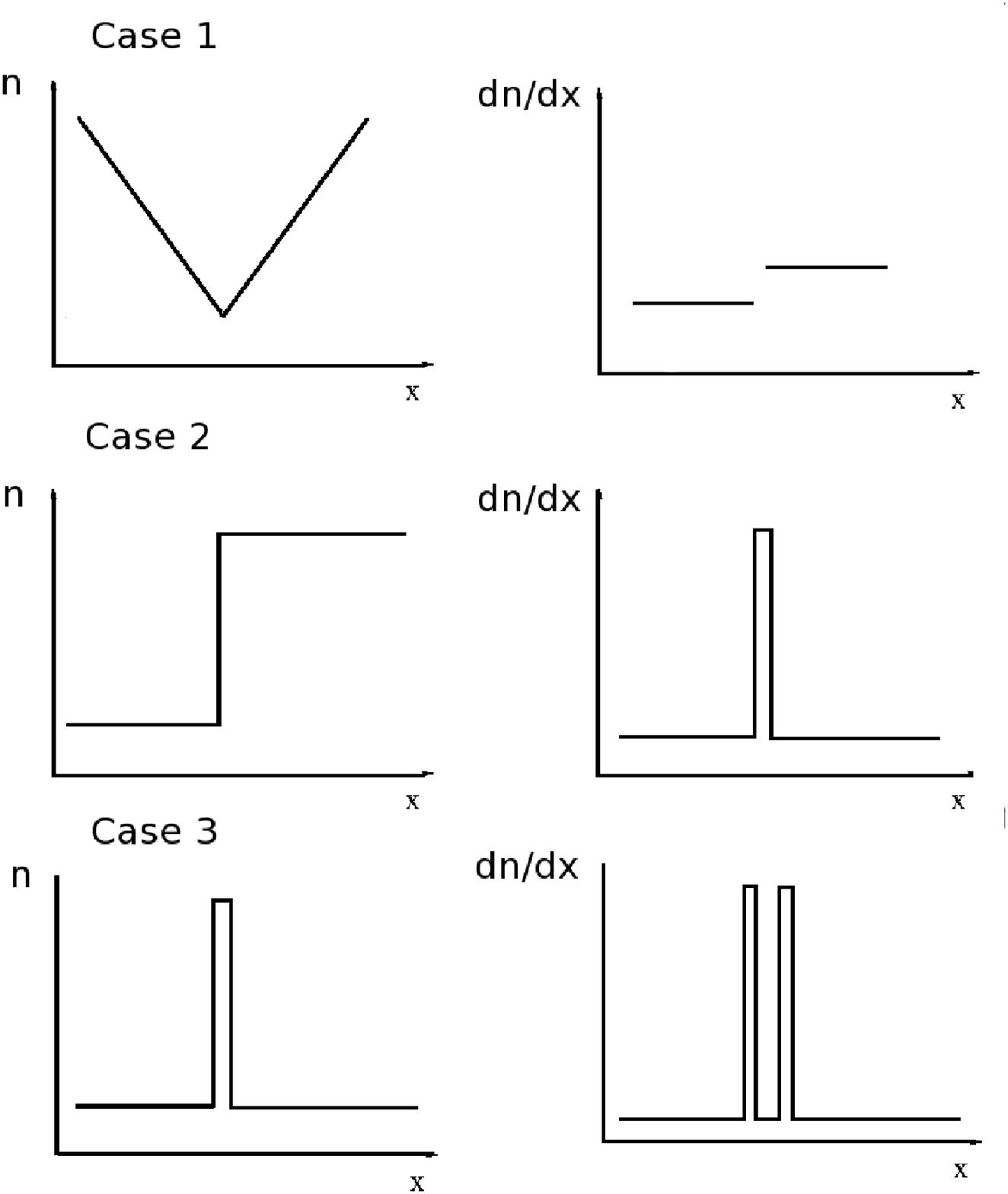}
\caption{Cartoon example of three possible scenarios for enhancements in a generic image ``n" , where  ``n" could be  $\textbf{P}|$, $RM$ or $\rho$ (density or EM).
Case one (top row) shows an example of 
a H\"{o}lder continuous function that is not differentiable at the origin (applicable to all turbulent fields).  Case two (middle row) shows an example
of a jump resulting from strong turbulent fluctuations along the LOS or weak shocks.
case three (bottom row) shows a delta function profile resulting from interactions of strong shocks.  In this case, 
the derivative gives a 'double jump' profile which produces morphology that is distinctly different from the previous cases.}
\label{fig:jumps}
\end{figure}

In the ISM, fluctuations in density and magnetic field will occur as a result of MHD turbulence,  which will be visible in polarimetric  maps.
In the case of taking gradients of a turbulent field,  one would expect to find filamentary structure
created by shock fronts, jumps and discontinuities.
Figure \ref{fig:jumps} shows a cartoon illustrating these three separate cases of a possible profile and its respective derivative. 
The cases are:
\begin{enumerate}
\item A H\"{o}lder continuous profile\footnote{H\"{o}lder continuous functions satisfy $|u(x_1, t)-u(x_2,t)|<C|x_1-x_2|^h$.  When the exponent h=1, this satisfies the Lipschitz condition.  
In the case of turbulent fields $ h=\frac{1}{3}$}  that is not differentiable at a given point (e.g. the absolute value function at the origin): Common for all types of MHD turbulence.
\item A jump profile: Weak shocks, strong fluctuations or edges (e.g. a cloud in the foreground which suddenly stops). 
\item A spike profile (e.g. delta function): Strong shock regime.
\end{enumerate}

In respect to case one, it is known that the turbulent velocity field in a Kolmogorov-type inertial range both
in hydro and  MHD are not differentiable, but only H\"older continuous (Bernard et al. 1998, Eyink 2009). 
Another example is that of any fractal function that displays self-similarity but is not differentiable everywhere.
This profile will naturally create discontinuities when one takes its derivative.  Therefore, case one can be found in both subsonic and supersonic type turbulence.
Case one type filaments can be seen in Figure \ref{fig:RM1}  for N, B, and $\textbf{P}$ in the right column.
Case two creates a structure in the gradient by a shock jump or a large fluctuation in either $n_e$ or B.  Here again, this type of enhancement in $|\nabla \textbf{P}|$ could be found in 
supersonic and subsonic type turbulence and is  due either to large random spatial increases or decreases due to turbulent fluctuations along the LOS or weak shocks.
We expect weak shock turbulence to show a lager amplitude in  $|\nabla \textbf{P}|$ then the subsonic case.
Case three is unique to supersonic turbulence in that it represents a very sharp spike in  $n_e$ and/or B across a shock front. The difference between this case
and what might be seen in case two is that here we are dealing with interactions of strong shock fronts, which are known to create delta function like distributions 
in density (Kim \& Ryu 2005).  In this case, the derivative of case three is has a distinctly different profile with respect to case one and two.  
Case three shows a 'double jump' profile across the shock front, which can be seen in Figure \ref{fig:RM2}
in the top right panel for LOS density and the bottom right panel for $|\nabla \textbf{P}|$.  
This morphological distinction can be used to determine if one is dealing with 
turbulence that is in a shock dominated regime (i.e. supersonic) and can  provide researchers with a promising new avenue of obtaining the sonic Mach numbers from 
polarimetric data.  In the case of the Alfv\'enic Mach number, the morphological difference is less clear, however gradients
will tend to align along the field lines in the case of strong field (sub-Alfv\'enic turbulence).

Also of interest is the question of which quantity is providing the dominate contribution to the structures in $|\nabla \textbf{P}|$ or $|\nabla RM|$: $\nabla n_e$, $\nabla B_{LOS}$
or both equally? Especially in the case of compressible turbulence, the  magnetic
energy is correlated with density, namely, denser
regions contain stronger magnetic fields, which is due to
the compressibility of the gas (Burkhart et al. 2009). This causes the magnetic field to follow
the flow of plasma if the magnetic  tension is negligible. The compressed
regions are dense enough to distort the magnetic
field lines, enhance the magnetic field intensity, and effectively
trap the magnetic energy due to the frozen-in condition.
Thus, for the supersonic cases, the intensity of the structures seen in $\nabla \textbf{P}$ are more pronounced then in the subsonic case, which is observed when
comparing Figures \ref{fig:RM1} and \ref{fig:RM2}.

However, in the case of subsonic turbulence, there are no compressive motions.  In this case, random fluctuations in density and magnetic field will create structures
in  $|\nabla \textbf{P}|$ and  $|\nabla RM|$.

\begin{figure}[h]
\centering
\includegraphics[scale=.5]{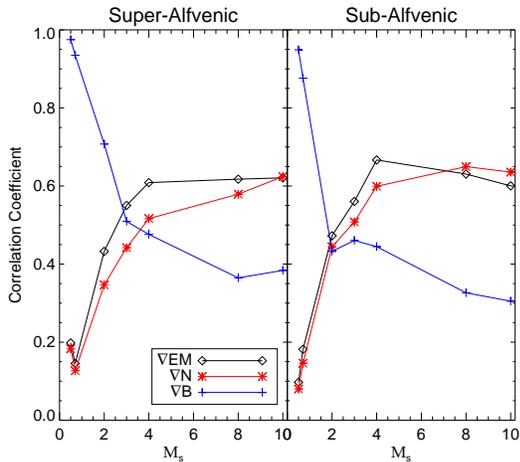}
\caption{Correlation coefficient between$|\nabla \textbf{P}|$ and the LOS $\nabla EM$ (black diamonds), $\nabla N$(red asterisk) and $\nabla B$ (blue plus sign).   
The left panel is super-Alfv\'enic and the right panel is sub-Alfv\'enic. In the case of strong shocks, strong correlation
is observed between $\nabla EM$ and  $\nabla N$ due to enhanced density. Subsonic models show strong correlations
with $\nabla B$. }
\label{fig:corav}
\end{figure}

Due to these effects, we might expect different trends in the correlation of supersonic and subsonic $|\nabla \textbf{P}|$ with 
$\nabla N$ or $\nabla EM$ (the gradient of the emission measure) and $\nabla B$.
We test this by plotting the pixel by pixel correlation coefficient of $|\nabla \textbf{P}|$ with the gradients of EM, N, and LOS B in Figure \ref{fig:corav}.  
In the case of subsonic turbulence,   $|\nabla \textbf{P}|$ better traces out the fluctuations in $\nabla B$ (blue line), while the supersonic cases
are dominated by density fluctuations.  This is because density enhancements are dominate due to shock fronts in the case of supersonic turbulence, while in 
subsonic turbulence density is highly incompressible.  In this case, the magnetic field will dominate the topology of the rotation measure and $|\nabla \textbf{P}|$.
This behavior is analogous to velocity in neutral hydrogen radio position-position-velocity cubes of turbulence, where density dominates the power spectrum for the case of supersonic
turbulence and velocity dominates the spectrum for subsonic turbulence (see Lazarian \& Pogosyan 2006, Burkhart et al. 2011a).
This difference in correlation provides yet another way of gauging the Mach numbers if one has both the emission measure
 and the linear polarization map.  Correlated spatial gradients between the two should indicate regions of shocks.

In the next section we will explore the utility of gradients of polarimetric data for the determination of the Mach numbers by investigating two different statistical 
measures of looking at the distribution and topology 
of the $|\nabla \textbf{P}|$ maps: PDF moments and genus function.

\section{Statistical Determination of the Sonic Mach Number}
\label{ms}
The previous section provided some theoretical discussion for why we expect $|\nabla \textbf{P}|$ data to be useful for determining the sonic Mach number.
In this section we will attempt to statistically quantify the differences seen in both the morphology and the distribution of maps of $|\nabla \textbf{P}|$.
We again note our assumption for the simulations of $|P|=1$ thus giving a trivial scaling relationship between  $|\nabla \textbf{P}|$ and $|\nabla RM|$ as:
$|\nabla RM|=|\nabla \textbf{P}|/2 $.  
We also provide an observational comparison for both statistics with the SGPS test region shown in Figure \ref{fig:RM}.

\subsection{Moments}
\label{pdfs}

A probability distribution function (PDF) is the function describing the frequency of occurrence of values in the distribution of intensities.  PDFs and their quantitative descriptors 
have been used to study turbulence in a variety of astrophysical context including diffuse ISM turbulence 
(Berkhuijsen \& Fletcher 2011), turbulence characterization (Federrath et al. 2010, Esquivel et al. 2010, Audit \& Hennebelle 2010)
solar wind (Burlaga et al. 2007) and molecular ISM (Padoan et al. 1999). Several authors have discussed the use of PDFs in determining
the Mach numbers of ISM turbulence, in particular the sonic Mach number (Vazquez-Semadeni 1994, Padoan et al. 1999,
Kowal, Lazarian \& Beresnyak 2007, Burkhart et al. 2010, Price, Federrath, \& Brunt 2011). However, this technique is almost always used in the context of the column density.  
To our knowledge, no one has applied this technique to the rotation measure, polarization maps, or their gradients.

One method of describing PDFs is by using statistical moments to characterize the mean and variance and departures from Gaussianity.

The first and second order statistical moments (mean and variance) used here are defined as follows:
$\mu_{\xi}=\frac{1}{N}\sum_{i=1}^N {\left( \xi_{i}\right)} $
and $ \nu_{\xi}= \frac{1}{N-1} \sum_{i=1}^N {\left( \xi_{i} - \overline{\xi}\right)}^2$, respectively. 

The 3rd and 4th order moments, Skewness and kurtosis respectively,  are defined as:

\begin{equation}
\gamma_{\xi} = \frac{1}{N} \sum_{i=1}^N{ \left( \frac{\xi_{i} - \overline{\xi}}{\sigma_{\xi}} \right)}^3 
\label{eq:skew}
\end{equation}

\begin{equation}
\beta_{\xi}=\frac{1}{N}\sum_{i=1}^N \left(\frac{\xi_{i}-\overline{\xi}}{\sigma_{\xi}}\right)^{4}-3
\label{eq:kurt}
\end{equation}

Where N is the total number of elements and $\xi_{i}$ is the distribution of intensities.

Past works have focused on the relationship between the moments and the density or column density.  As the Mach number increases, so does the mean value of density,
as shocks increase density (the sonic Mach number goes as ${\cal M}_s \approx \rho ^{1/2}$).  The variance, skewness, and kurtosis
are less obvious quantities and their relation to the sonic Mach number has been derived numerically (Padoan et al. 1999, Kowal, Lazarian \& Beresnyak 2007, Burkhart et al. 2010
Price, Federrath, \& Brunt 2011).  Variance has some disadvantages to skewness and kurtosis.  One is that
variance is scale dependent, and values will change between different data set normalizations.  This makes direct comparison between simulations and 
observations difficult. On the other hand, the higher order moments (skewness and kurtosis) describe deviations from Gaussianity and are unit-less numbers, and therefore
are scale free.  They also are shown to increase more linearly with the sonic Mach number in the case of column density (Kowal, Lazarian \& Beresnyak 2007, Burkhart et al. 2010).  
However, the increase is not so pronounced in the case of subsonic turbulence, making variance a better indicator in this regime. 
\begin{figure}[h]
\centering
\includegraphics[scale=.6]{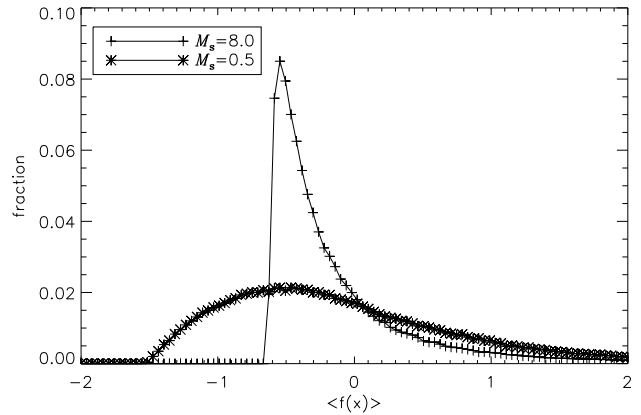}
\caption{PDFs of $|\nabla \textbf{P}|$ normalized as $f(x)=(|\nabla \textbf{P}|-<|\nabla \textbf{P}|>)/\sigma(|\nabla \textbf{P}|)$ where $\sigma(|\nabla \textbf{P}|)$ 
is the standard deviation of $|\nabla \textbf{P}|$. 
The supersonic case (represented by + signs) is highly skewed and kurtotic compared with the subsonic case (represented by * signs).}
\label{fig:PDF}
\end{figure}

While the rotation measure is obviously related to the density, the relationship between the moments  of  $|\nabla \textbf{P}|$ and the Mach number has never been studied.  
The polarization gradient may have an advantage over column density/dispersion measure for investigating the Mach numbers of the WIM, as shocks will cause both an increase in 
density and magnetic field due to correlations between density and field strength in supersonic type turbulence (see Burkhart et al. 2009).  Furthermore, the gradient will highlight interesting
features in the observational data which might be buried by observational effects, such as a DC offset, that otherwise would not be seen in the maps of 
column density or linear polarization (see Figures \ref{fig:RM1} and \ref{fig:RM2}).  

\begin{figure}[htb]
\centering
\includegraphics[scale=.6]{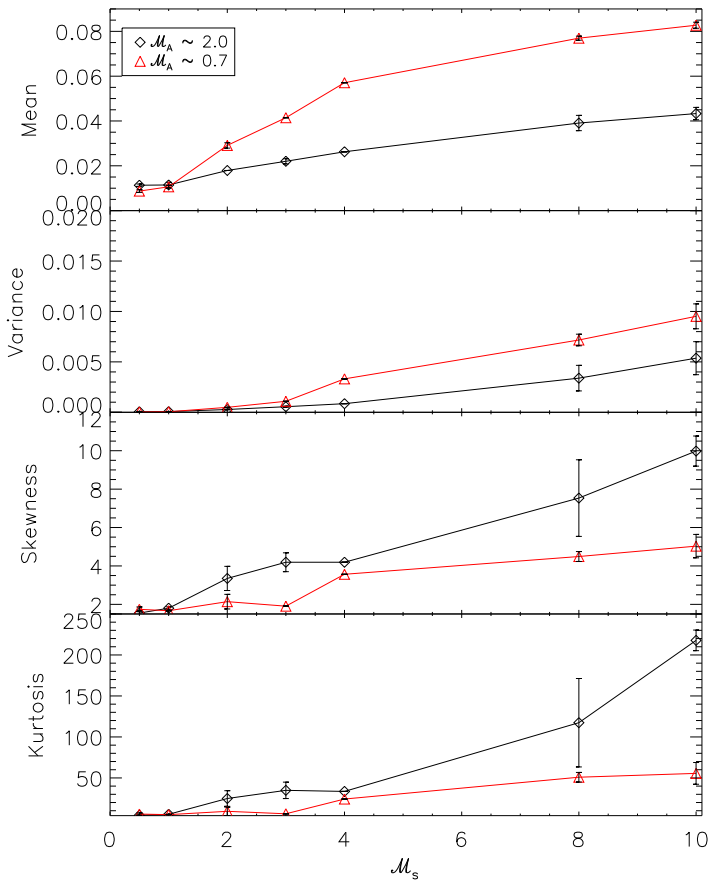}
\caption{Moments of  $|\nabla \textbf{P}|$ vs. ${\cal M}_s$.  Error bars are created by taking the 
standard deviation of the moments between different time snapshots of the well-developed turbulence.   We show sub-Alfv\'enic cases in red and super-Alfv\'enic cases in black. }
\label{fig:moments}
\end{figure}

Figure \ref{fig:PDF} shows an example PDF of a subsonic and supersonic simulation (model 1 and 6 ).  Here it is clear that the general relations that are true
for the column density will apply here: as the Mach number increases the distribution becomes more kurtotic and more skewed.  
Figure~\ref{fig:moments} shows the moments of  $|\nabla \textbf{P}|$ vs. the sonic Mach number for our simulations with LOS taken along the mean magnetic field.
Error bars are created by taking the 
standard deviation between different time snapshots of the data.  We see that all four moments tend to increase with increasing sonic Mach number.
This is due to the shock fronts creating more discontinuities and sharper gradients. This not only increases the mean value and the variations in 
$|\nabla \textbf{P}|$, but also creates very peaked distributions and distributions with tails skewed towards the right.  

\begin{figure}[htb]
\centering
\includegraphics[scale=.45]{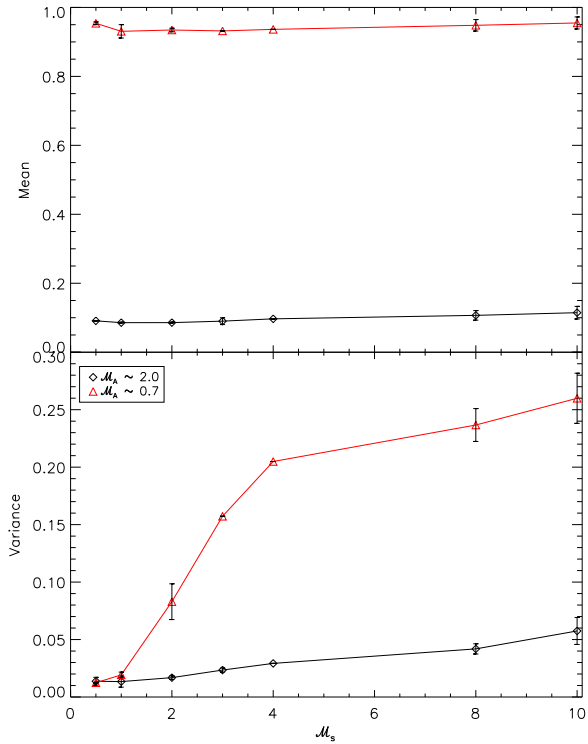}
\caption{Mean (top) and variance (bottom) of the polarization angle (or RM$\lambda^2$) \textit{along the direction of the mean magnetic field} vs. sonic Mach number.  A strong
Alfv\'enic dependency is observed.  In the case of the mean value, the dependency is independent of the sonic Mach number.  The trend is different for a different LOS. }
\label{fig:moments_rm}
\end{figure}

In terms of the Alfv\'enic Mach number, the main contributor is the LOS direction chosen relative to the ordered field.  When we look parallel to the mean field
as shown in in Figure~\ref{fig:moments}, the sub-Alfv\'enic cases show higher mean and variance while the super-Alfv\'enic cases show generally higher skewness and kurtosis.
We plot the mean and variance of the \textit{polarization angle} in Figure \ref{fig:moments_rm}, which confirms the trend seen in the gradients.  In the case of 
the LOS parallel to the mean field, the mean value of the polarization angle has a strong Alfv\'enic dependency with no sonic Mach number dependency.
The variance shows strong dependency on both sonic and Alfv\'enic Mach numbers.  However, these trends are unique for sight-lines parallel to the mean field LOS.
When looking perpendicular to the LOS the effects change.  For example,   the moments of  $|\nabla \textbf{P}|$ show stronger skewness and kurtosis in the case
of the sub-Alfv\'enic models.  Thus to gauge the Alfv\'enic Mach number using the distribution, the ordered LOS magnetic field is required.

\begin{figure*}[htb]
\centering
\includegraphics[scale=.6]{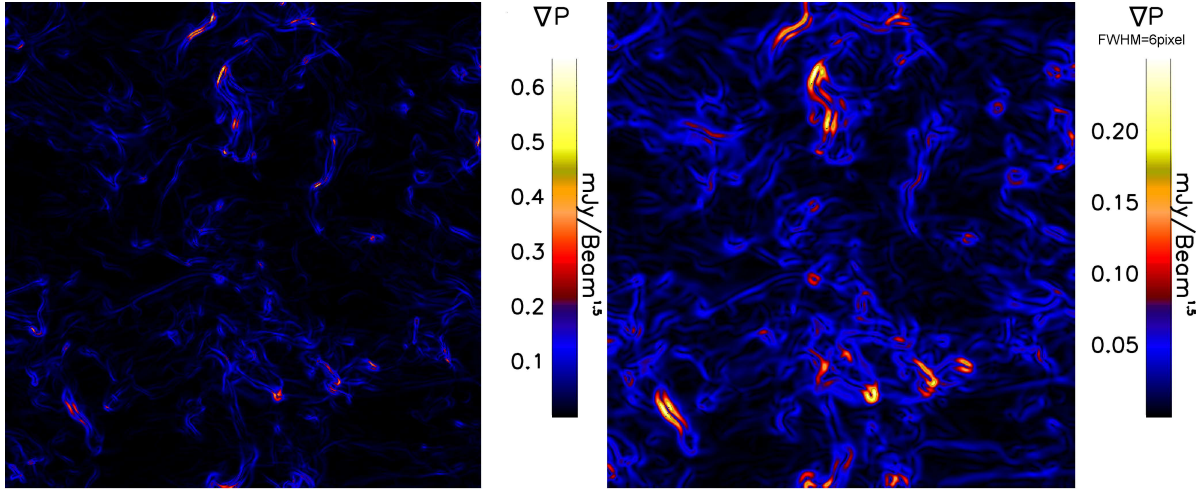}
\caption{Maps of $|\nabla \textbf{P}|$ for model 6  with smoothing FWHM=6 pixels (right) and without smoothing (left). The telescope resolution has a strong effect on the distribution of  $|\nabla \textbf{P}|$  
yet the morphology remains similar.}
\label{fig:smoothiimage}
\end{figure*}

\begin{figure}[htb]
\centering
\includegraphics[scale=.6]{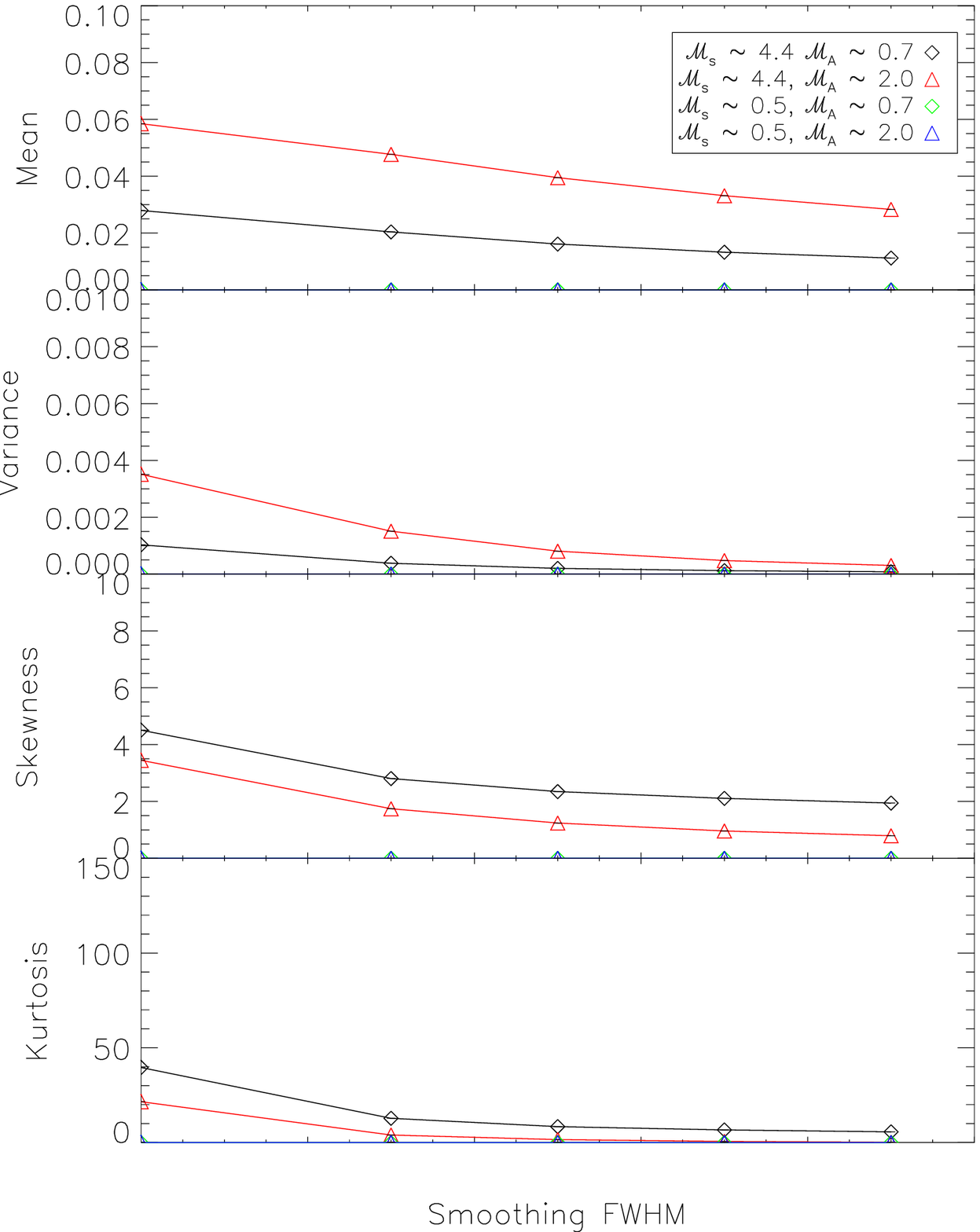}
\caption{Moments of  $|\nabla \textbf{P}|$ vs. smoothing  for four different models. Error bars are created by taking the 
standard deviation between different time snapshots of the data.   Smoothing FWHM is given in pixels.}
\label{fig:momentsm}
\end{figure}

An additional effect that must be considered is the issue of the telescope resolution.  For example, Figure \ref{fig:smoothiimage} shows model number 6 
without smoothing (left) and with Gaussian smoothing with FWHM=6 pixels (right).  It is clear that smoothing changes the distribution of 
maps of $|\nabla \textbf{P}|$.
We plot the moments vs. smoothing FWHM for  four models in Figure \ref{fig:momentsm}.  As the resolution of maps of Q and U decrease, so do the moments.
Thus one needs to take the smoothing of the data into account when comparing PDFs of  $|\nabla \textbf{P}|$.

We investigate the moments of regions of the  SGPS data with the most emission removing the bad pixels from our investigation.  We find 
that the SGPS data has average  skewness of  0.825  and kurtosis of 0.928 and mean value and variance of 0.004 and $3.38x10^{-6}$, respectively.
Comparing with our numerical set up, this most closely matches subsonic to transonic values of the moments, with telescope smoothing taken into account.

\subsubsection{Moment Maps}
\label{mommap}
From section 4.1 it is clear that,  for  a given map of  $|\nabla \textbf{P}|$, as the sonic  Mach number increases the skewness and kurtosis also increase.
However, these were globally averaged values of both the moments (i.e. the PDF of the entire image) and the sonic Mach number. 
If we want to look at smaller scale variations, we can calculate the PDFs of smaller portions of the image. 
Using a moving kernel (box, circle etc.), we can create a ``moment map" which is essentially a smoothed map
that calculates the moments at every point with a given box size.  Because we are dealing with gradient quantities, we expect
these maps to particularly trace regions where shocks are interacting and along shock fronts, thus tracing areas
where the sonic Mach number is \textit{changing}.

We make  skewness and kurtosis moment maps of the $|\nabla \textbf{P}|$  images similar to the method of Burkhart et al. (2010). 
This results in a smoothed moment image,  which can be compared to the actual image of the LOS local sonic Mach number (LOS average
sonic Mach number for each pixel) with the same smoothing kernel as the $|\nabla \textbf{P}|$ moment map.
In this section, we investigate the relationship between the LOS sonic Mach number and the values of skewness and kurtosis on a pixel-by-pixel basis. 

The key questions one must consider when making a moment map are what the kernel shape should be,
 what resolution is appropriate for good statistics, and how to compare the values of the higher order moments to
the sonic Mach number? The last point is particularly important.  While Burkhart et al. (2010) took a linear fit between the sonic
Mach number and the moments in the case of column density,  we will
test how effective the moment maps can distinguish between supersonic and subsonic regimes on a pixel-by-pixel bases.  
Thus we stress that we are not trying to calculate the exact sonic Mach number with this method, which is a topic for another work.  We only
seek to determine how well a moment map is able to pick out regimes of supersonic and subsonic.
For our initial test, we choose to use a boxcar kernel for both the $|\nabla \textbf{P}|$ maps and for averaging the sonic Mach number,
since a boxcar  can handle edges in a square image most effectively. 

How well the moment map is able to determine if the gas is supersonic or subsonic depends on three parameters:
the threshold level for skewness  ($\gamma_T$, above which the gas is considered supersonic and below it is considered subsonic),
 the threshold level for kurtosis ($\beta_T$, above which the gas is considered supersonic and below it is considered subsonic), and the kernel box size chosen. 
 The values for  $\gamma_T$ and $\beta_T$ represent the threshold value between supersonic and subsonic regimes.  Any pixels  below either $\gamma_T$ and $\beta_T$  is deemed subsonic and 
any pixel above both is deemed supersonic.  Of course, an intermediate threshold value could also be chosen to probe the transsonic regime, but we omit this here. 
Again we should stress that we are not employing this method to get an exact Mach number;
 we simply view it as a way of determining whether our $|\nabla \textbf{P}|$  image shows supersonic or subsonic characteristics.

The best fitting parameter values of box size, $\gamma_T$ and $\beta_T$ are those that provide accurate translation between the moments and the LOS Mach number.  In order to determine these values
we use a genetic algorithm which searches possible combinations of these three parameters in order to determine the best fit.  The genetic algorithm can provide more computationally efficient way of 
testing for fitness rather then iterating over every possible combination of parameter space.  See the appendix for a detailed description of our algorithm.

\begin{figure}[htb]
\centering
\includegraphics[scale=.6]{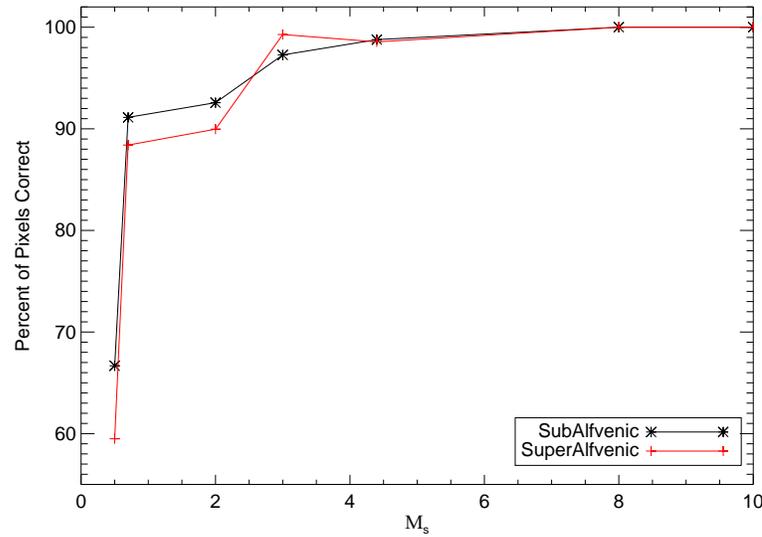}
\caption{The success rate of the moment map.  A moment map is constructed with parameters $\gamma_T$=1.1 and $\beta_T$=1.58 and box size (64 pixel), which were
chosen with a genetic algorithm.
If a value in the moment map is above both $\gamma_T$ and $\beta_T$, then that pixel is considered supersonic.  If the value is below either, then
the pixel is considered subsonic.  
The moment map is compared with the actual LOS sonic Mach number map to determine whether the moment map alone is sufficient for recovering regimes of
supersonic or subsonic on a pixel-by-pixel basis. The moment map is very successful (almost 100\%) in determining if supersonic turbulence is present.
The success rate drops to $\approx$60\% in the case of diagnosing subsonic turbulence. This further confirms the higher order moments utility in the presence of supersonic
flows.}
\label{fig:corprct}
\end{figure}

\begin{figure}[htb]
\centering
\includegraphics[scale=.6]{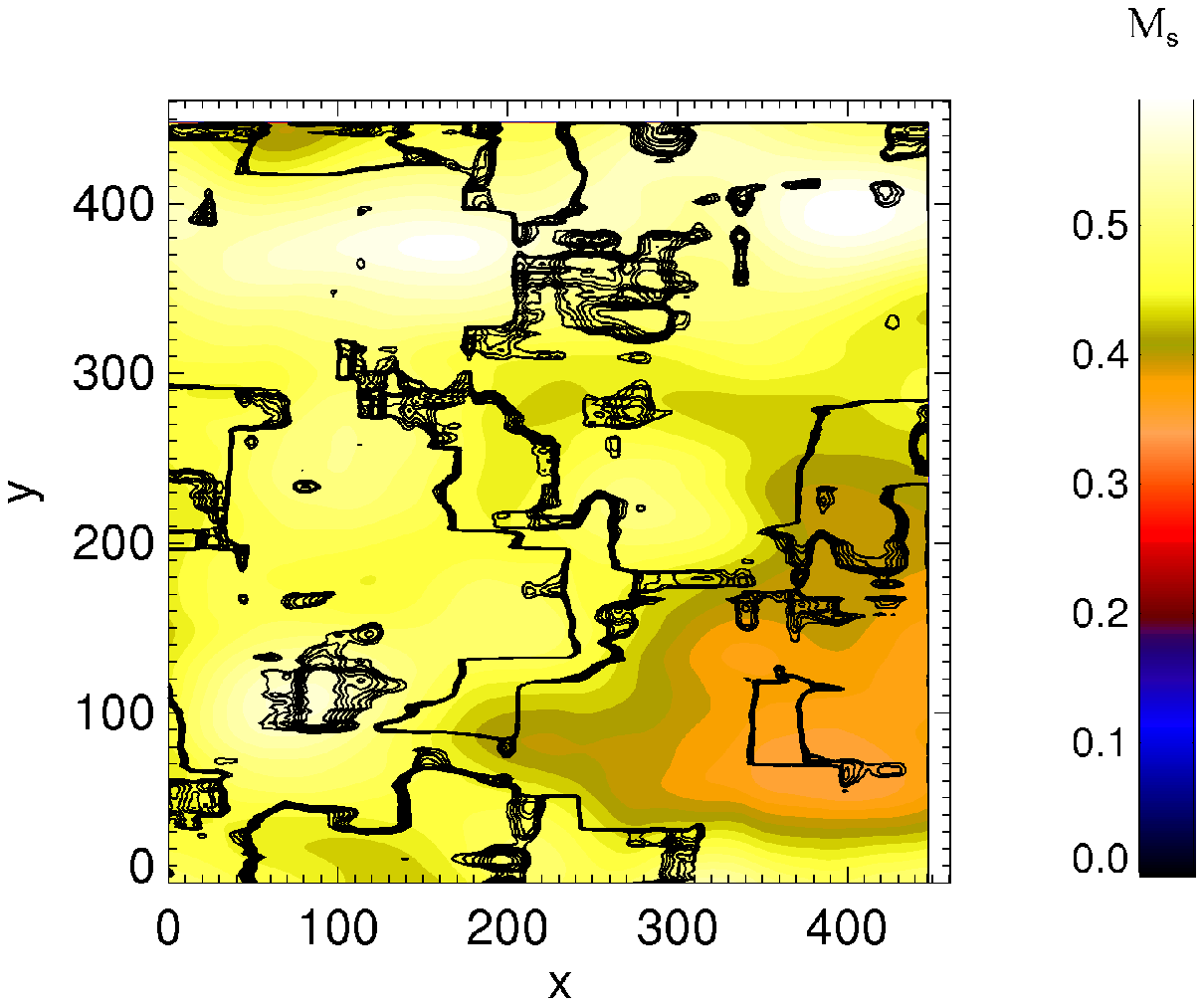}
\includegraphics[scale=.6]{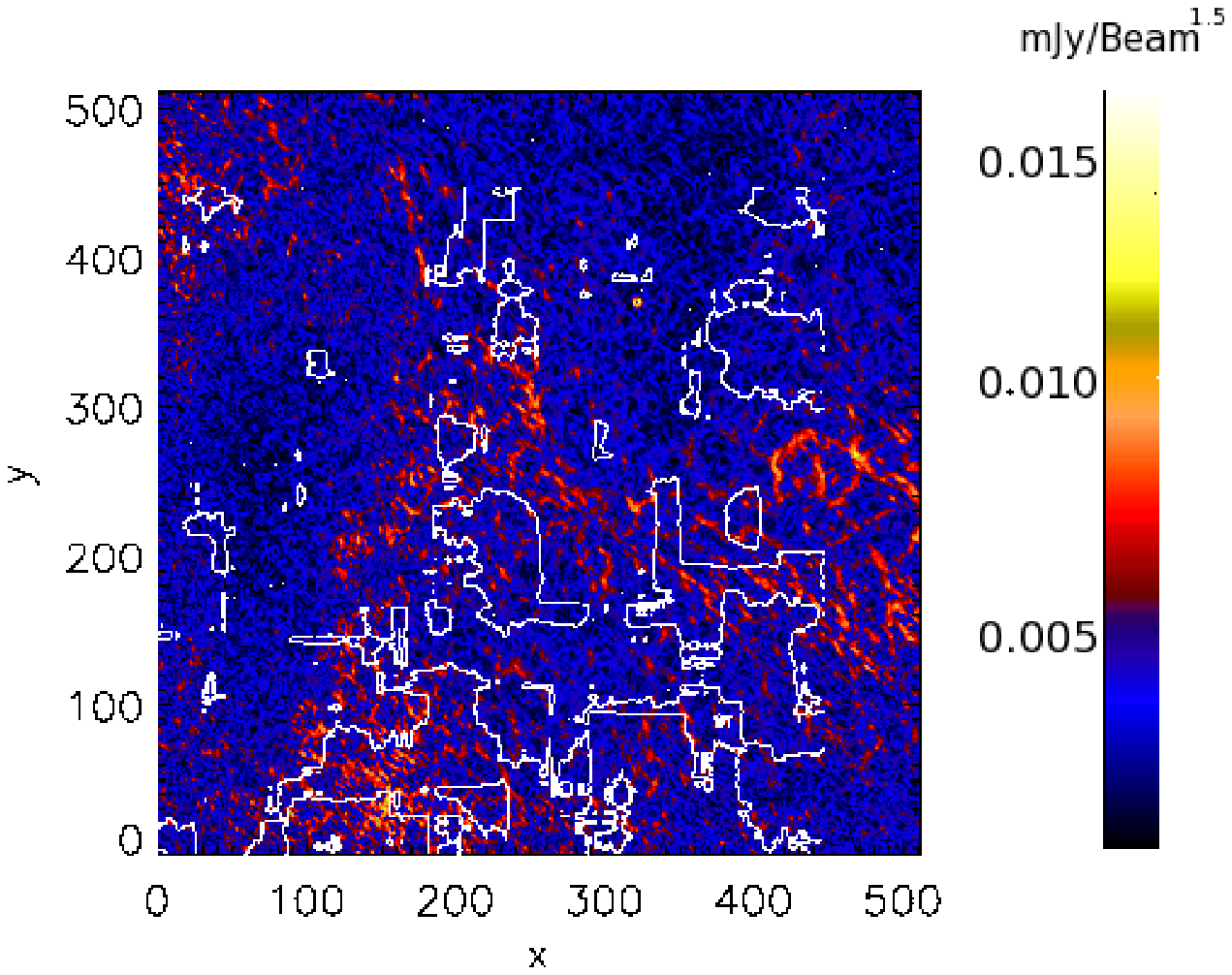}
\caption{Top: LOS sonic Mach number map for model number 1 with over-plotted contours of the kurtosis moment map with $\gamma_T$=1.1 and $\beta_T$=1.58 and smoothing
kernel= 64 pixels. The moment map generally outlines areas where the sonic Mach number is changing, which is expected for a gradient quantity. 
Bottom: Map of the SGPS test region used in the paper with the kurtosis moment map contours over-plotted.}
\label{fig:mommap_obs}
\end{figure}

Although multiple combinations of parameters showed high confidence levels, we chose  $\gamma_T$=1.1 and $\beta_T$=1.58 and box size=64 pixels.
With these values, the supersonic models were able to determine the Mach number regime with 98\% accuracy, while the subsonic cases had an accuracy of 67\%
(see Figure \ref{fig:corprct}).  This is not surprising however, since
the moments are known to be a more robust measure of highly supersonic turbulence  (Kowal, Lazarian \& Beresnyak 2007).  
We plot  the LOS Mach number map for model number 1 in the top panel of Figure \ref{fig:mommap_obs} with contours from the kurtosis moment map.
The moments trace areas where the sonic Mach number is changing.  

We apply the moving box method to the SGPS  data cuts using the same parameters for  $\gamma_T$, $\beta_T$  and box size as were used for the simulations.
For the SGPS data  we obtained average skewness and kurtosis values for this moment map of 0.3 and  0.9 for skewness and kurtosis respectively.   We over-plot the SGPS data
with the kurtosis moment map contours in the bottom panel of Figure \ref{fig:mommap_obs}.   
While these values seem to point in the direction that the SGPS data is subsonic or transonic, we note that this method is less accurate for subsonic 
type turbulence.  However, it does confirm that the gas in this patch of the sky is not statistically similar to what is expected for supersonic flows. 


\subsection{Topology: The Genus Statistic}
\label{genus}

The filaments seen in $|\nabla \textbf{P}|$  show substantially different morphology when comparing maps of subsonic and supersonic turbulence, as was discussed in 
Section 3.  Thus, a natural
avenue of characterization would be to use topological measures in order to pick out different structures.  In this section we will investigate the utility of the
genus statistics in order to characterize the topology of $|\nabla \textbf{P}|$  filaments. 

The genus statistics was developed to study the topology  and deviations from Gaussianity of the
universe and the distribution of galaxies in three dimensions (Gott
et al. 1986; 1987). The use of genus statistics for the study of HI was first 
discussed in Lazarian (1999), and subsequent studies presented the genus curves for the SMC (Lazarian et al. 2002;
Lazarian 2004, Chepurnov et al. 2008) and for MHD simulations (Kowal, Lazarian \& Beresnyak 2007).

Genus is a quantitative measure of topology. It can characterize
both 2D and 3D distributions. 
Generally speaking, the genus is used to detect departures for Gaussianity.
When dealing with the  ISM  one cannot expect deviations from symmetry to be
small, especially in the presence of supersonic flows. In this case, genus can be used to 
characterize flows that are supersonic since these  show large deviations from Gaussianity.

 The 2D genus can be represented
as (Coles 1988; Melott et al. 1989):
\begin{align*}
 G \equiv \text{(\# isolated high-density regions)}  & \\ - \text{ (\# isolated low-density regions)} 
 \end{align*}
where low- and high-density regions are selected with respect to
a given contour threshold.  
For instance, a uniform circle would have a genus of 0 (one
connected region of high density, i.e., an ``island,'' and one
connected region of low density, while a ring (a donut, for example)
would have a genus of -1 (one
connected region of high density and two connected regions
of low density).   Thus the  genus can distinguish between
``meatball" and ``swiss cheese"  topologies (Gott 1990).
\begin{figure}[tbh]
\centering
\includegraphics[scale=.5]{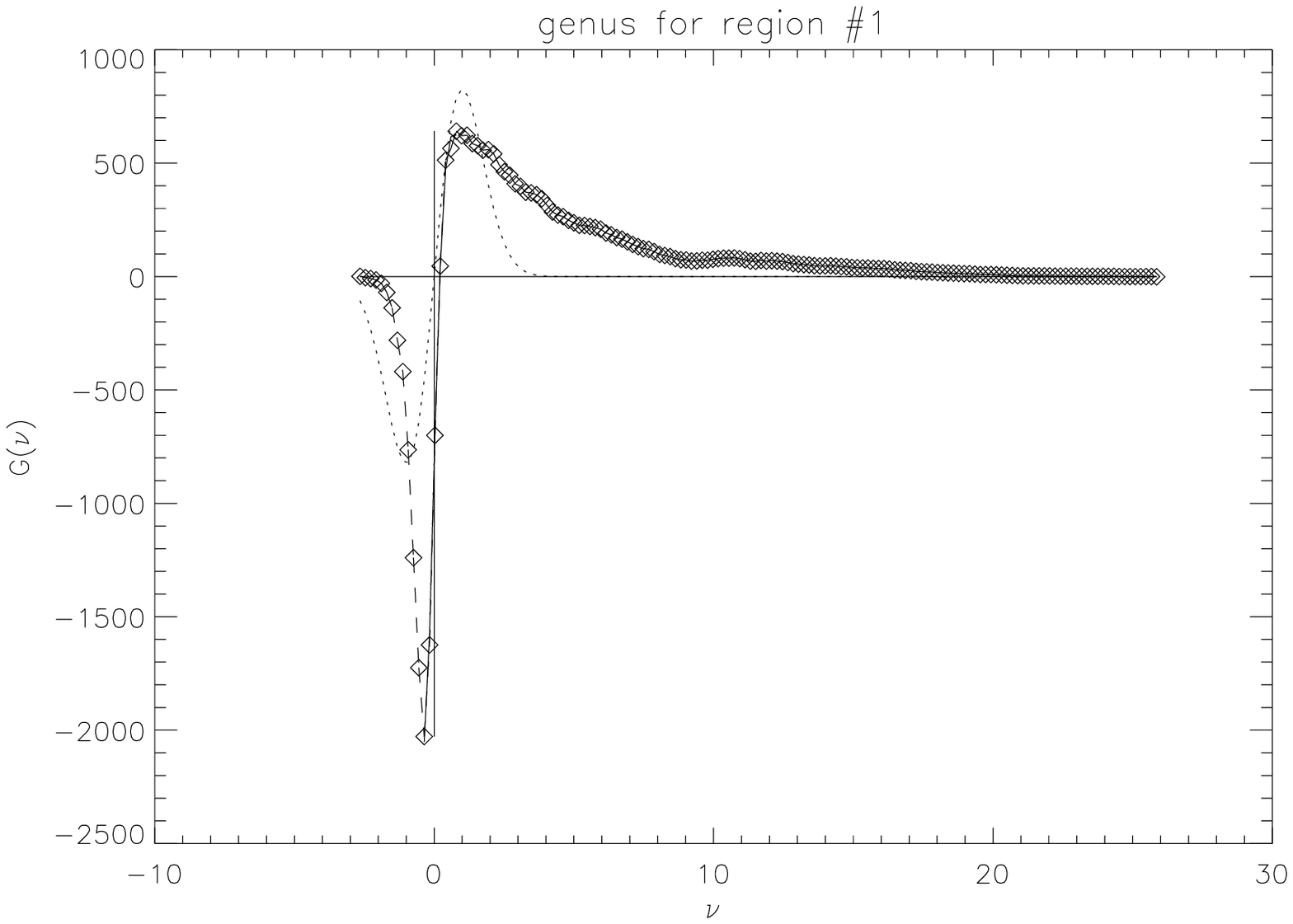}
\caption{Genus curve of  $|\nabla \textbf{P}|$  for  model number 1 (pictured in the bottom right of  Figure \ref{fig:RM}).  
The genus curve of a Gaussian distribution is shown in dashed lines.
The $|\nabla \textbf{P}|$ genus deviates  strongly from the Gaussian model.   Horizontal and vertical solid lines reference the origin.}
\label{fig:genus}
\end{figure}

For a 2D image, the genus is simply a number which corresponds to 
a given threshold value.  What is considered to be high or low is dependent on the
threshold value, which acts as a free parameter. As a result, for a given 2D image,
the threshold value can be varied to construct a curve with the genus value on the y-axis and the threshold value on the x-axis. 
This is known as the genus curve.
We show an example genus curve for  $|\nabla \textbf{P}|$  in Figure \ref{fig:genus}.  This genus curve is for the subsonic sub-Aflv\'enic (model number 1) simulation.  
In the case of a density or column density field, the genus of subsonic turbulence is close to a Gaussian field.  However
the $|\nabla \textbf{P}|$ distribution has longer tails and a more pronounced minimum.
The maximum and minimum points of the genus curve correspond to
percolation of the distribution (see Colombi et al. 2000). The fact
that the observed genus falls more slowly at large thresholds than 
 the Gaussian distribution indicates that the
islands are more discrete and pronounced than for the Gaussian
distribution.

\begin{figure}[tbh]
\centering
\includegraphics[scale=.5]{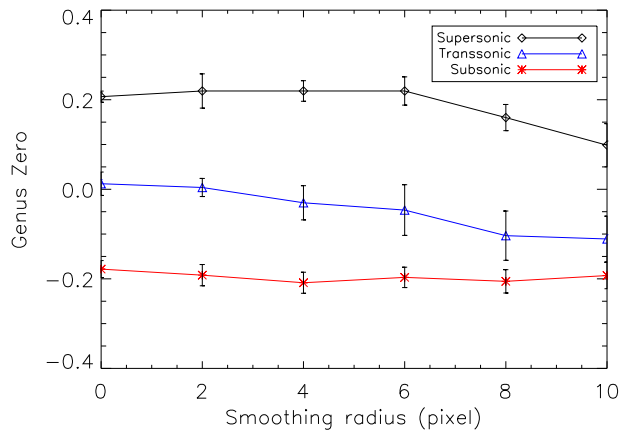}
\caption{Genus zero values of simulated $|\nabla \textbf{P}|$ .  Negative values imply the topology is clumpy (meatball) while positive values imply
the topology is hole dominated (swisscheese). The genus of $|\nabla \textbf{P}|$ is fairly insensitive to smoothing of maps of  Q and U.
The supersonic cases show a hole topology  while the subsonic case shows a clump topology, even in the case of smoothing. Transonic cases show 
topology that is a mixture of clumpy and hole dominated.}
\label{fig:gradcd}
\end{figure}

We expect that the sign of the genus curve at the mean intensity level of  $|\nabla \textbf{P}|$ (i.e. where the curve crosses the x-axis)
 does describe the field topology.
A positive genus
will represent a clump-dominated field, while a negative one will
mean the domination of holes. However, it is more convenient to
work with the zero of the genus curve (genus zero) $\nu_0$, because it can be normalized to the field variance. 
In other words, consider a map in which the mean value of the map is subtracted off.   $\nu_0$ represents
the location where the genus curve crosses the x-axis, with the origin being the mean value of the field, so that $G(\nu_0)=0$.
In this case, for the intensity with the subtracted mean value, a negative $\nu_0$ corresponds
to the clumpy topology, while a positive $\nu_0$ indicates the ``swiss cheese" topology.

\begin{figure}[tbh]
\centering
\includegraphics[scale=.5]{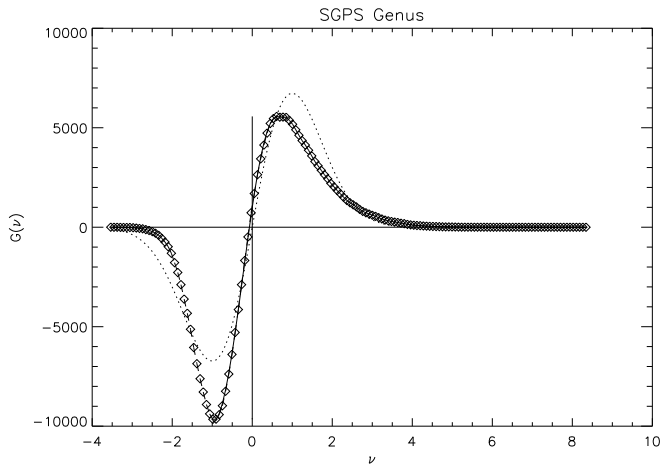}
\caption{Genus curve for $|\nabla \textbf{P}|$ of the SGPS data set.  The genus zero is at -0.07 which indicates a slightly clumpy topology. This is most similar
to the transonic type genus curves (${\cal M}_s \approx 2.0$).  Horizontal and vertical solid lines reference the origin.}
\label{fig:sgps_genus}
\end{figure}

We make plots of the genus zero  vs. smoothing in Figure \ref{fig:gradcd}.  A negative shift indicates a clumpy topology, while a positive shift indicates a hole topology.  
The error bars are derived by estimating the variance of the genus distribution.  For the error bar creation, we follow the method used by Chepurnov et al. (2008)
in that we generated  a set of images
with randomly shifted phases of individual harmonics. This procedure causes the field to take Gaussian statistics, and therefore
approximately the genus zero is at the origin.  However, slight deviations from Gaussian  allows us to effectively estimate its variance.
The procedure is as follows: we take a fast Fourier transform (FFT) of
the region being studied and assign the phase of each harmonic
to a random variable uniformly distributed.   After the inverse
FFT, we calculate the respective $\nu_0$.
After repeating this procedure ten times, we calculate the variance of the $\nu_0$ values. 

Figure \ref{fig:gradcd} plots the genus zero vs. smoothing because the telescope resolution is an important
quantity to consider in any statistical analysis of observational data.  
As was demonstrated with the PDF moments in the previous subsection, as maps of Q and U are smoothed, the distribution tends towards Gaussian.
As such we include analysis of the genus zero with smoothing to see what effect the telescope resolution has on the topology. 
We include the genus zero vs. smoothing plots  for maps of $|\nabla \textbf{P}|$ in the  supersonic, transsonic and subsonic turbulence regimes. 
We see that over a range of  smoothing values
the topology between the three regimes  is very different.   The supersonic case shows a positive genus zero indicating that the topology is more hole like.  
The topology of the subsonic case is more clump like.   Visual inspection of Figure \ref{fig:RM1} confirms this.  
In the case of the transonic turbulence, the topology is more neutral.  Interestingly, there is not a strong dependency when we 
smooth the maps of Q and U.  The topology remains roughly consistent over a range of smoothing.

We show the genus curve for the SGPS $|\nabla \textbf{P}|$ data in Figure \ref{fig:sgps_genus}.  The genus zero is at -0.07, which indicates a slightly clumpy topology.  
This value more closely matches the transsonic values accounting also for smoothing effects.




\section{Discussion and Conclusions}
\label{concl}
Quantitative analysis (via PDF moments and genus) of the polarization gradient indicates that  turbulence in the 
ionized ISM in these directions is in the range of subsonic to transonic. Our findings are supported by recent studies
of Balmer-$\alpha$ emission measures, which have similarly found
Mach number of 2 or less (see Hill et al. 2008).  For studies of neutral gas, i.e.  21 cm HI gas, the warm phase of the SMC was shown to have 
properties of transonic gas (Burkhart et al. 2010).  In the case of the cold component, several independent  statistical
and direct measurements applied to the spectral lines and the column density indicate this component of the SMC is supersonic (Stanimirovic \& Lazarian 2001, Burkhart et al. 2010). 
Furthermore, a recent study by Chepurnov \& Lazarian (2010) added the Wisconsin H-$\alpha$ Mapper (WHAM) emission measure data to the 
 ``Big Power Law" of electron density fluctuations (see Armstrong et al. 1995)
provided further proof for a parsec to AU -5/3 power law, indicating that this phase of the ISM is not highly compressible.  
Thus it becomes apparent that different types of data, sampling different phases of the ISM, 
can compliment each other and provide more clues towards furthering our understanding of a turbulent ISM.

Observational studies of ISM turbulence are extremely important. From the observational data one would like to obtain the characteristics of turbulence, e.g.
 intensity of its driving, importance of magnetic field in the dynamics of the media. These characteristics can be evaluated if we know sonic and Alfv\'en Mach numbers, i.e.
$ {\cal M}_s$ and ${\cal M}_A$. There have been several attempts to develop the techniques to get these numbers from observations using both column density data  (see Kowal, Lazarian \& Beresnyak 2007  Burkhart et al. 2009)
 as well as spectroscopic data (Burkhart et al. 2011a). An example of successful application of such techniques to the Small Magellanic Cloud (SMC) HI data is provided in Burkhart et al. (2010).

This paper explores the utility of a new observational motivated technique, namely the gradient of the polarization  map,  to determine ${\cal M}_s$ and ${\cal M}_A$. 
The observational advantage of using gradients stems from the fact that these gradients are easily available from interferometric observations. Our study shows that $|\nabla \textbf{P}|$ is a very useful 
measure which allows one to study turbulence.  We might also expect gradients be useful for turbulence studies 
in other types of data sets where shock morphology will be observed (i.e. column density, see Figures \ref{fig:RM1} and \ref{fig:RM2}).

We view this paper as a first taste of the utility of using polarization data (and their gradients) for studies of turbulence.  There are many additional avenues
that should be explored from this first step.  For instance, what are the effects of multiple screens along the LOS?  What are the effects of changing the assumption
of the constant background polarization?  What are the effects of other equations of state and the inclusion of partial ionization?  

One may wonder whether an additional way of studying turbulence is valuable, if we already have a few other ways to study turbulence, e.g. with column densities and position-position-velocity 
(PPV) spectroscopic data cubes. The answer is a sounding yes!. First of all, dealing with as complex media as the ISM, we would like to have as many independent measures as possible.
Second, different measures may be more sensitive to different phases of the ISM (see the list of the idealized phases and their magnetizations in Yan, Lazarian \& Draine 2004). For example, the 
rotation measure, linear polarization and their gradients are biased towards ionized parts of ISM.

The Alfv\'en and sonic Mach number are not the only characteristics of the ISM turbulence. 
Spectra of turbulence, its intermittencies (see Kowal \& Lazarian 2010) provide other measures which should
 be used to study interstellar turbulence in its complexity.

\subsection{Conclusions} 

We created maps of the spatial gradient of the polarization vector of isothermal MHD simulations  and compared these with observations.
We tested two statistical methods on gradient polarization maps, namely the genus and higher order moments of the distribution to determine if these
statistics were sensitive to the sonic Mach number.   We found that:
\begin{itemize}
\item Filamentary structure was created over a range of sonic Mach numbers, including cases where both shocks and subsonic turbulence were present. 
\item Filaments showed different morphology for different regimes of sonic Mach number.
\item $|\nabla \textbf{P}|$ maps with high sonic Mach number showed filaments with a double jump profile 
that traced shocks, while subsonic cases showed filaments that were due to random fluctuations in the rotation measure along the LOS.  Transonic simulations had both
types of filaments present
\item The moments of the $|\nabla \textbf{P}|$ distribution were higher for larger values of sonic Mach number but were also sensitive to the telescope resolution.
\item There is a strong Alfv\'enic dependency in the moments of polarization angle and $|\nabla \textbf{P}|$ for sight-lines parallel to the ordered magnetic field. 
\item The skewness and kurtosis moment maps of $|\nabla \textbf{P}|$ were successful at picking out subsonic or supersonic pixels  67\% and 99\% of the time, respectively.
\item The genus of $|\nabla \textbf{P}|$ revealed a ``hole"  topology for supersonic cases and a ``clump"  topology for subsonic cases.  Transsonic cases showed neutral
topology.  This trend is generally smoothing independent.  
\item We applied the PDF moments and genus to the  SGPS test region and found that this area was statistically similar to models of subsonic to transsonic type turbulence.
\end{itemize}

B.B. acknowledges support from the NSF Graduate Research Fellowship and the NASA Wisconsin Space Grant
Institution. B.B. thanks Andrew Schechtman-Rook  for valuable discussions.
 A.L. thanks both NSF AST 0808118 and the Center for Magnetic Self-Organization in Astrophysical and Laboratory Plasmas for financial support. 
This work was completed  during the stay of A.L. as
Alexander-von-Humboldt-Preistr\"ager at the Ruhr-University Bochum and the University of Cologne.
B.M.G. acknowledge the support of an Australian Laureate Fellowship awarded by the Australian 
Research Council through grant FL100100114. The Australia Telescope Compact Array is funded
 by the Commonwealth of Australia for operation as a National Facility managed by CSIRO.
\appendix

\section{The Genetic Algorithm}
We employ a genetic algorithm to search for the optimal three parameters ($\gamma_T$ and $\beta_T$ and box size) in the creation of the moment map, as discussed in Section 4.
The literature on the genetic algorithm is vast, so we refer the curious reader to Eiben \& Smith (2003)  and references therein.
Unlike brute force methods, the genetic algorithm relies on principles of biological evolution, such as reproduction and natural selection, to determine the optimal parameters. 

A typical genetic algorithm requires
a fitness function to evaluate the solution domain.  In our case the fitness metric is how well the local skewness and kurtosis are able to determine if the local sonic Mach number
is either in the supersonic or subsonic regime. In our case, local means on a pixel-by-pixel basis.  The main advantage of us choosing a genetic algorithm over a brute force method is the cost in computation time.  Rather then looping over all possible
combinations of parameters, we only seek a convergence to a range of the best fit outcomes.  We ran the algorithm several separate times on multiple simulations
with varying sonic and Alf\'venic Mach numbers in order to assure that the convergence obtained was repeatable.

The algorithm we employ is a simple implementations of a genetic algorithm and works as follows.  We start with a population of box sizes and threshold values.  The algorithm initially picks a random subset of
 ten  candidates from this larger population pool; that is, it picks ten combinations of $\gamma_T$, $\beta_T$, and box size. Then the moment map is created by calculating the moments in the kernel, then moving
 the kernel pixel-by-pixel and repeating the moment calculation.  

 We take this map and compare it to the LOS sonic Mach number map (smoothed by the same degree at the moment map) pixel-by-pixel. 
  If a pixel in the moment map is above $\gamma_T$ and $\beta_T$ then it is considered supersonic, else it is subsonic.
 If this matches with what is seen in the LOS sonic Mach number map then we give this pixel a value of 1, if it does not match then we give this pixel a value of 0.
 We calculate the percentage of successful pixels
to determine the fitness of that particular choice of parameters.

Once we have determined the percent of success for all ten of these initial candidates, we ``clone" the most fit of these and then ``breed" the candidates to fill out the other nine models of the next generation. 
This keeps the total population size constant.  Our breeding is done by averaging the parameters
from a random selection of these ten candidates (the parent population).
The most successful candidate from the parent population and the averaged candidates now become the ``children population."
The algorithm is repeated  from the point of making the moment map.  This process continues for 10 iterations of parent and child.  
Because our algorithm uses small population sizes, after a fairly small number of generations the variety in free parameters will shrink
significantly, resulting in an ``inbred" population. Therefore every 10th generation, instead of breeding the next generation, we clone the
fittest model and then fill out the rest of the subsequent generation with new models, randomly chosen in the same way as the models used
in the first generation of the algorithm. This process ensures that the genetic algorithm always has a large region of parameter space to explore.

 We ran the genetic algorithm multiple times to ensure the solution of best fit was converging to roughly the same parameters. While there were many  parameter sets in the 90\%+ success range in the supersonic case, 
 we choose  $\gamma_T$=1.1 and $\beta_T$=1.58 and box size=64 pixels, as it was the best match we found with the subsonic cases (67\%)






\end{document}